\documentclass{elsart}

\def\simgt{\lower 2pt \hbox{$\, \buildrel {\scriptstyle >}\over {\scriptstyle \sim}\,$}}
\def\simlt{\lower 2pt \hbox{$\, \buildrel {\scriptstyle <}\over {\scriptstyle \sim}\,$}}

\gdef\kms{km\,s$^{-1}$}

\begin{document}

\begin{frontmatter}

\title{Observing Cluster Galaxies and their Progenitors with JDEM}

\author{Pieter G.\ van Dokkum}
\address{Department of Astronomy, Yale University, New Haven,
CT 06520-8101}

\begin{abstract}

The Joint Dark Energy Mission (JDEM) is expected to have a
field-of-view that is several orders of magnitude larger than that of
current instruments on the Hubble Space Telescope,
with only slightly reduced sensitivity and
resolution.  This contribution gives a brief discussion of the
impact that JDEM would have on studies of galaxies in clusters
at $0.5<z<2$ and of the most massive galaxies at higher redshift.
Of particular importance is JDEMs unique wide-field
near-IR capability, enabling
the selection and morphological study of high redshift galaxies
in the rest-frame optical rather than the rest-frame ultra-violet.

\end{abstract}

\end{frontmatter}

\section{Introduction}

The primary goal of the Joint Dark Energy Mission (JDEM) is to improve
our understanding of the nature of dark energy through accurate
brightness measurements of Type Ia supernovae (SNe) out to $z\approx 1.7$.
However, JDEM could have a profound impact on many other fields of
astronomy:
if it is designed as a wide-field
optical/near-IR survey telescope, whose full-resolution data are
saved and sent to Earth, it will be a tremendous improvement over
current facilities and an ideal complement to the ``pencil-beam''
JWST. Compared to the current capabilities of HST, JDEM
would have a field-of-view $100 \times$ greater in the optical
and $1500 \times$ greater in the near-infrared, with only slightly
lower sensitivity and resolution. The
Deep Survey envisioned by the SNAP collaboration (Aldering et
al.\ 2004) would image an
area of 15 degrees$^2$ in 9 optical/near-IR filters to greater depth
than the Hubble Ultra Deep Field.

Many applications of JDEM were highlighted at the meeting, including
the impact it will have on the studies of large scale structure (see
contribution by Daniel Eisenstein), galaxy formation (Rachel
Somerville), and galaxy evolution (David Hogg). This contribution
focuses on the importance of JDEM for the study of galaxy clusters at
$0.5<z<2$ and of the likely progenitors of cluster galaxies at higher
redshift. A SNAP-like design is assumed, i.e., a wide
field optical/near-IR imager.

Studies of the formation and evolution of cluster galaxies give
insight in the assembly and star formation history of the most massive
disk- and spheroidal
galaxies in the Universe. Because of their high masses and old
stellar populations they
provide important tests of the hierarchical paradigm for galaxy
formation: in these models galaxies are built up slowly through a
combination of mergers and star formation, and more massive galaxies
are predicted to have
assembled more recently (e.g., Meza et al.\ 2003).
The evolution of
cluster galaxies is also important for understanding galaxies at very
high redshifts. In hierarchical models the first objects form
preferentially in overdense regions, which evolve into groups
and clusters at the present day. Specifically, Baugh et al.\ (1998)
predict that the descendants of $z\approx 3$ Lyman break galaxies
(LBGs) typically live in halos of circular velocity $> 400$\,\kms\ today,
compared to $\sim 250$\,\kms\ for galaxies that did not have a LBG
progenitor. These arguments also apply to the descendants of
the brightest galaxies at $z\sim 7$ and beyond, and to the
highest redshift quasars.


\section{What are we learning from HST?}

As so many other fields, the study of cluster galaxies
has benefited greatly from the combination of HST
imaging and spectroscopy with 8--10m ground-based
telescopes in the past decade.
At the time of the meeting, about 60 distant
clusters had been observed with WFPC2 and about 30 with ACS
(excluding snapshot programs).
The clusters
were selected in X-rays
or in the optical/near-IR, and span a redshift range
$0.2\simlt z \simlt 1.4$.
The vast majority of the HST observations cover
just a single central pointing of $<10$ arcmin$^2$, and only a handful
studies have observed areas larger than the equivalent of $\approx 4$
independent pointings (e.g., van Dokkum et al.\ 1998a,
Treu et al.\ 2003).

\subsection{Evolution of early-type galaxies}

Early-types (elliptical and S0 galaxies) constitute $\sim 80$\,\%
of the galaxy population in the cores of nearby clusters
(Dressler 1980). They
form a very homogeneous population: at a given mass,
they show a very small scatter in their colors, luminosities,
and line indices. This high degree
of regularity is expressed in tight scaling relations between
color and magnitude (Bower et al.\ 1992), velocity dispersion
and Mg line strength (e.g., Bender et al.\ 1993), and velocity
dispersion, radius, and surface brightness (the Fundamental
Plane; Djorgovski \& Davis 1987).

Studies of the redshift evolution of these relations are in remarkable
agreement. The evolution of their colors (e.g., Ellis et al.\ 1997,
Stanford et al.\ 1998, Blakeslee et al.\ 2003), $M/L$ ratios and
Fundamental Plane (e.g., van Dokkum et al.\ 1998b,
Kelson et al.\ 2000, van Dokkum \& Stanford 2003, Wuyts et al.\ 2004),
and line indices (Bender et al.\ 1998,
Kelson et al.\ 2001) show that massive early-type galaxies remain
a very homogeneous
population over the entire redshift range $0<z<1.3$.

The strongest constraints on the mean age of the stars in early-type
galaxies have come from studies of the Fundamental Plane (FP).
Figure 1a shows the evolution of the mean $M/L_B$ ratio of early-type
galaxies with masses $>10^{11}\,M_{\odot}$, as determined from the
FP (van Dokkum et al.\ in prep.). The evolution is
well described by the passive fading of a stellar population formed
at $z\approx 2.8$ (indicated by the line), with very small
cluster-to-cluster scatter. The best constraints on the galaxy-galaxy
scatter in the ages have come from HST studies of the color-magnitude
relation. By combining data from the literature with
new ACS images of a cluster at $z=1.24$ Blakeslee et al.\ (2003) find that
the scatter in rest-frame $U-B$ colors is only a few percent
independent of redshift (Fig.\ 1b).

\begin{figure}
\includegraphics{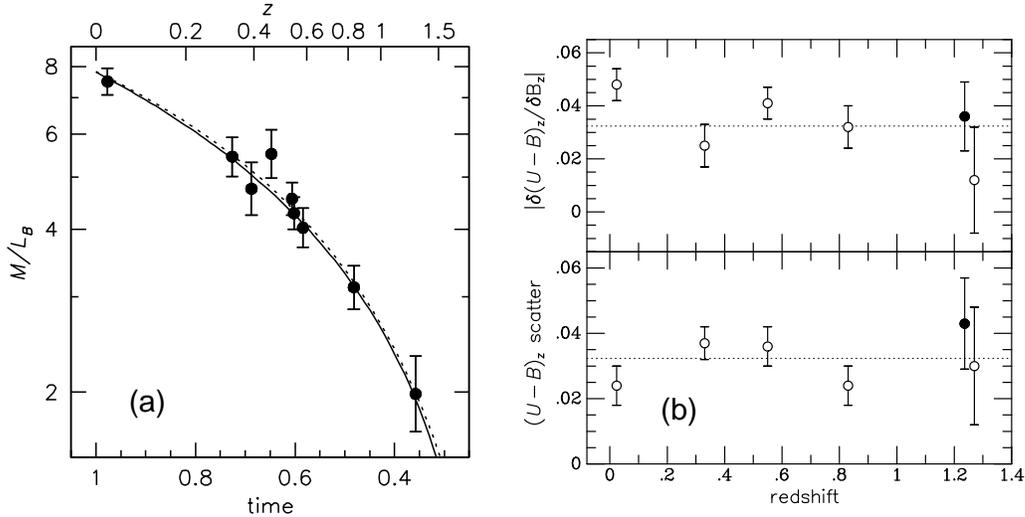}
\vspace{6.7cm}
\caption{Early-type galaxies in the cores of rich clusters appear
to evolve slowly and regularly. The evolution of their
$M/L_B$ ratios
is consistent with passive fading of stellar
populations formed at $z\approx 3$
(panel a; van Dokkum et al.\ 1998b and in prep.), and there is no evidence
for evolution in the slope or scatter of their color-magnitude
relation (panel b; Blakeslee et al.\ 2003).
}
\end{figure}

\subsection{Morphological evolution: interaction with the environment}

It has been known for a long time that this simple picture of early
formation and passive evolution cannot be complete. The earliest
evidence for significant recent evolution in cluster environments
was the discovery of the Butcher-Oemler effect: the increase with
redshift of the fraction of blue galaxies in clusters (e.g.,
Butcher \& Oemler 1978, Ellingson et al.\ 2001, de Propris et al.\ 2003).
Furthermore, HST studies of the central regions of
clusters have demonstrated a decreasing fraction of early-type
galaxies with redshift (Dressler et al.\ 1997; van Dokkum
\& Franx 2001; Smith et al.\ 2004): the early-type fraction in clusters at
$z\sim 1$ is $\approx 50$\,\%, compared to $\approx 80$\,\% in the
local universe. This trend is qualitatively consistent with
the larger fraction of blue galaxies at higher redshift, although
there is no one-to-one relation between morphology and color,
in particular at bright magnitudes
(e.g., Poggianti et al.\ 1999,
de Propris et al.\ 2003). Also, galaxies have been
``caught in the act'' of transforming their morphology and/or
spectral type. Examples are ``E+A'' galaxies, which
have early-type spectra with
strong Balmer lines indicating a recent star burst;
red merger systems found in several 
clusters at $z\sim 1$ (e.g., van Dokkum et al.\ 1999);
and galaxies being stripped of their cold gas
in the Virgo cluster (e.g., Kenney et al.\ 2004).

These effects are probably at least in part driven by infall of
galaxies from the field. Clusters are expected to accrete a
significant fraction of their final mass after their initial collapse
(e.g., Diaferio et al.\ 2001).  Simulations and
observations suggest that the infalling galaxies are likely to
radically change
their star formation rate and morphology as a result of
interactions with other galaxies and the hot X-ray gas (e.g., Abraham
et al.\ 1996, Moore et al.\ 1996, Abadi et al.\ 1999, Poggianti
et al.\ 1999, Ellingson et
al.\ 2001, Kodama et al.\ 2001). The details of these
processes are not well understood, largely because of the difficulty
of studying clusters out to the virial radius and beyond (i.e.,
to $R\sim 10$\,Mpc).
Treu et al.\ (2003) showed that
with a large investment of HST time radial trends in morphology can be
established out to the virial radius. However, as demonstrated
in Fig.\ 2 HST is very inefficient in covering such large areas,
in particular in the near-infrared.


Whatever the cause, morphological transformations severely
complicate the interpretation of studies
of galaxy evolution in clusters. First of all, it means that
we cannot equate the mean age of the stars in a galaxy to its
``assembly age'', the time since the galaxy took on its current
appearance. Furthermore, if
many present-day early-types were recently transformed from
(field) spiral galaxies
their progenitors are not included in samples of $z\sim 1$
cluster early-type galaxies. This ``progenitor bias''
may cause us to underestimate the evolution of early-type galaxies
in clusters (van Dokkum \& Franx 2001). The effects of this
bias can be quantified by determining the mechanism and rate
of morphological transformations in clusters, or by
interpreting the observed evolution of cluster galaxies in
the context of a full cosmological simulation (e.g.,
Diaferio et al.\ 2001). Both approaches are ambitious, and
require a much better understanding of physical processes in the
infall regions.

\section{The role of JDEM}

\subsection{The infall regions of clusters}


The large field-of-view of JDEM will enable simultaneous study of
clusters and the large scale structure in which they are embedded. The
build-up of clusters through infall of galaxies and groups along
filaments can be studied in detail, and the morphological changes that
are thought to occur in the infall regions can be quantified.
A SNAP-like JDEM design
will observe the entire field shown in Fig.\ 2 in a single
exposure, and owing to its unique near-IR capability
can extend wide-field studies of clusters to
$z\sim 2$. Among specific questions that JDEM
would address are the morphological composition of galaxies
in filaments and infalling groups; the merger rate as a function
of local density and dynamical state; and the prevalence
of ``passive spirals''
(e.g., Goto et al.\ 2003). The broader aims are to determine the rate
and mechanisms of morphological transformations as a function of
redshift.


\begin{figure}
\includegraphics{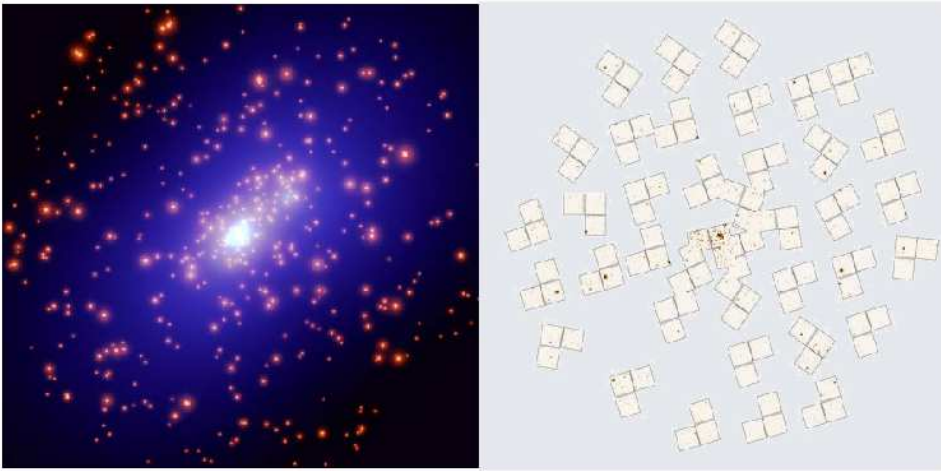}
\vspace{6.7cm}
\caption{Wide angle views of the cluster
CL\,0024+16 at $z=0.39$.
{\it Left:} mass map, with dark matter shown in blue and luminous
matter in red (Kneib et al.\ 2003). {\it Right:} 39 HST WFPC2 pointings sparsely
sample a $\sim 27'$ diameter field (Treu et al.\ 2003). Imaging such large
areas with HST is inefficient in the optical, and
virtually impossible in the near-infrared.
\vspace{0.3cm}
}
\end{figure}

A cluster survey could target specific clusters (selected in X-rays,
or by their Sunyaev-Zel'dovich effect), but could also piggyback
on planned SNe searches. Extrapolating from
lower redshifts one may expect $>10$ rich clusters in SNAP's
Deep Survey and $>500$ in its Wide-field Survey, depending
on the imposed mass and redshift cutoff. 
The clusters can be selected from the weak lensing maps produced
as part of the survey (Wittman et al.\ 2001, Aldering et al.\ 2004);
from X-ray surveys that will
likely be planned in the Deep Survey area (see
the contribution by Neil Brandt); and
with the Gladders \& Yee (2000) red sequence technique,
which is probably the most efficient method beyond $z\sim 1$.
The detailed weak lensing maps are not only useful for
selecting clusters, but
also characterize the dark matter substructure of clusters
and their surroundings
at $0.5\simlt z\simlt 1.2$ (Hoekstra et al.\ 2000).

The extreme depth of the Deep Survey also opens up the interesting
possibility of seeing the ``tidal tracks'' left by infalling
galaxies. Simulations suggest that clusters should be riddled
with ultra-low surface brightness tidal debris (e.g., Dubinski 1999),
providing a fossil record of the orbits of galaxies similar
to that provided by stars in the Galactic halo. It may also
be possible to see faint tidal features around a large fraction
of elliptical galaxies, if they were assembled at relatively
low redshift through ``dry mergers''\footnote{Rachel Somerville's
term.} involving little gas
(van Dokkum et al.\ 1999; Bell et al.\ 2004).

\subsection{Progenitors at $z>2$}

Galaxy overdensities have been identified out to $z\sim 4$ and beyond,
by selecting galaxies in narrowband
filters centered on the redshifted Ly$\alpha$ line
(e.g., Venemans et al.\ 2002) or by the
Lyman break technique (Steidel et al.\ 1998). However, normal
galaxies in the nearby universe and known
cluster galaxies out to $z\sim 1.5$ would not be selected in this way, as
they are too faint in the rest-frame ultraviolet.
The top panels of Fig.\ 4 illustrate the effect of selecting galaxies
in the UV: the GALEX view of M31 only shows the OB associations,
missing most of the stellar mass.

The bias toward unobscured star forming galaxies inherent in the Lyman
break technique can be avoided by selecting galaxies in the rest-frame
optical rather than the rest-frame UV, using their redshifted Balmer-
or 4000\,\AA-break. This selection is difficult due to the relative
weakness of the Balmer break and the high sky brightness in the
near-IR, and has only recently become feasible with the
advent of high quality, large format near-IR detectors on large
telescopes. Using very deep VLT images with a total area of
$\approx 30$\,arcmin$^2$ we recently found a large population of
galaxies at $2<z<3$ whose rest-frame optical
colors are much redder than those
of LBGs (Franx et al.\ 2003;
van Dokkum et al.\ 2003; F\"orster Schreiber et al.\ 2004). The
galaxies are efficiently selected by the simple observed color
criterion $J-K_s>2.3$.
They are very massive (up to $\sim 5 \times 10^{11}\,M_{\odot}$;
van Dokkum et al.\ 2004) and appear to be
highly clustered (Daddi et al.\ 2003). Their colors, H$\alpha$ equivalent
widths, line widhts, stellar masses, and metallicities are similar
to massive star forming galaxies in the nearby universe
(van Dokkum et al.\ 2004; F\"orster Schreiber et al.\ 2004; Fig.\ 3).
Most are
bright in the near-IR but too faint in the observer's optical to be
selected by the Lyman break technique. Their masses,
clustering, and colors strongly suggest that they are progenitors of
today's early-type galaxies.

\begin{figure}
\includegraphics{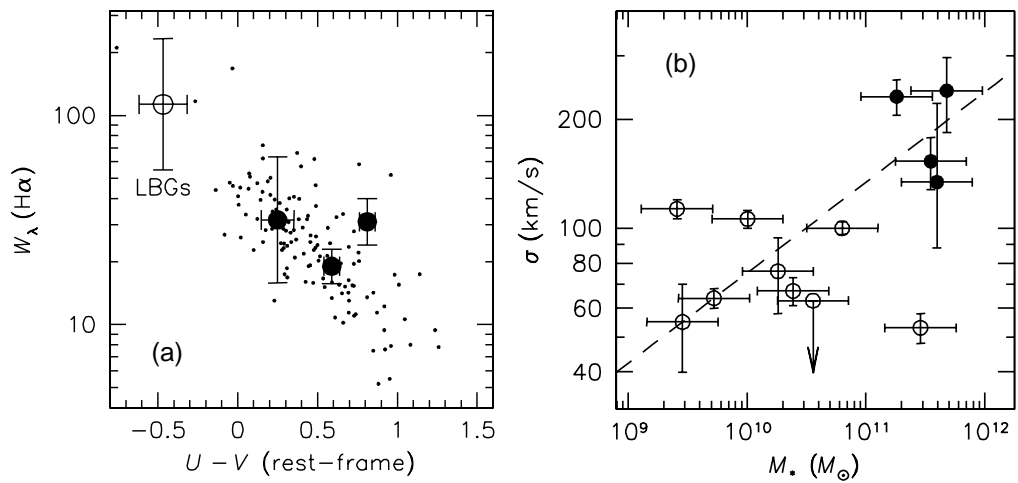}
\vspace{6.7cm}
\caption{Properties of $z\sim 3$ galaxies selected in the rest-frame
UV (LBGs/open circles; from Pettini et al.\ 2001 and
Shapley et al.\ 2001) and $z\sim 2.5$ galaxies selected by their red
rest-frame optical colors (solid circles).
The small sample of red $z\sim 2.5$ galaxies has
lower H$\alpha$ equivalent widths,
higher stellar masses, and higher line widths than typical UV-selected
$z\sim 3$ galaxies (from van Dokkum et al.\ 2004).\vspace{0.3cm}
}
\end{figure}

The strong field-to-field variations induced by their clustering
(see, e.g., Labb\'e et al.\ 2003) imply that large areas need to
be imaged to determine accurate surface densities and correlation functions.
However, deep, large area near-IR surveys are notoriously difficult.
Using $\sim 200$ hours of VLT observing time FIRES surveyed
an area of only 30\,arcmin$^2$ for red $z>2$ galaxies
(Franx et al.\ 2003; F\"orster Schreiber et al.\ 2004). Although
we are
currently extending this area in the MUSYC project\footnote{Multi-wavelength
Survey by Yale-Chile; www.astro.yale.edu/MUSYC/}, there is no
telescope/instrument configuration on the horizon that can
compete with JDEM in the near-IR.
In addition to exquisite photometry  JDEM will provide
morphologies in the rest-frame UV and optical of all detected
$z>2$ galaxies. The bottom panels of Fig.\ 4 illustrate the importance
of having multi-wavelength morphological information
for any $z>2$ galaxy which has a mix
of young and old stars. This near-IR selected galaxy in the UDF has an
irregular morphology in the optical ACS image, but is regular and
concentrated in the near-IR NICMOS image (Toft et al., in prep).
SNAP's Deep Survey would cover a $\sim 5000 \times$ larger area and go
$\sim 3$\,mag deeper in the near-IR than the UDF NICMOS campaign
(Thompson et al., in prep), at comparable resolution. If stellar
mass correlates with line width at $z>2$ (as indicated
by current data; see Fig.\ 3b), it may be possible
to determine the distribution function of
halo circular velocities of galaxies at $z\sim 2.5$
from JDEM photometry alone.

\begin{figure}
\includegraphics{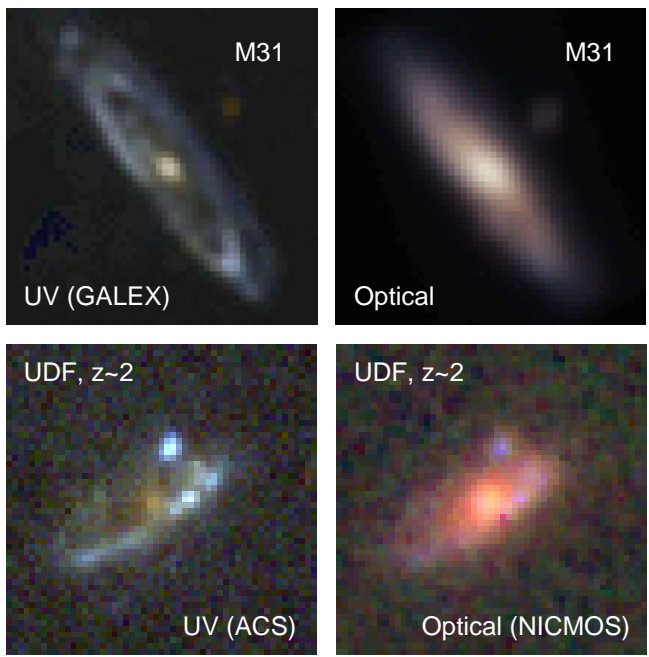}
\vspace{9.5cm}
\caption{Comparison of morphology in the rest-frame UV
and the rest-frame
optical, for M31 (top panels) and a red galaxy at $z\approx 2$
in the HST Ultra Deep Field (bottom panels; Toft et al., in prep).
The images have the same
resolution and physical scale ($40 \times 40$\,kpc), but arbitrary
relative intensity scaling. Galaxies whose stars
ghave a range of ages can look very different in the
rest-frame UV and the rest-frame optical. For such objects the
rest-frame UV will miss most of the stellar mass.\vspace{0.3cm}
}
\end{figure}

\subsection{JDEM design considerations}

JDEM will obviously be optimized for the identification and follow-up
of high redshift SNe. Nevertheless, it is interesting to consider how
well the current design specifications are suited for the study of
high redshift galaxies. In terms of resolution the current SNAP design
is similar to the Wide Field camera of WFPC2 in the optical and
the NIC3 camera in the near-IR. As demonstrated in Fig.\ 4, NIC3's
resolution is sufficient to determine the morphologies of normal
spiral and elliptical galaxies at arbitrary redshift. However, for
resolving bulges, dwarf galaxies,
and individual star forming complexes at $z\simgt
0.5$
one would have to sacrifice area for smaller ($\sim 0.10''$) pixels
in the near-IR.

The reddest JDEM filters determine the
highest redshifts at which the Lyman- and Balmer-breaks can be isolated.
In the current design SNAPs reddest filters are at 1.24\,$\mu$
and 1.44\,$\mu$, implying a maximum redshift for measuring the Balmer
break of $\sim 2.4$, and a maximum Lyman ``dropout'' redshift of $9-10$.
Given the potential of JDEM as a survey companion to JWST the addition
of a redder filter at $\sim 1.6\,\mu$ or beyond would be highly valuable.

Taking spectra of JDEM sources
will be a formidable challenge, in particular
in the near-infrared. As
even the spectroscopic limit of a 30m ground-based
telescope will be $5-6$
magnitudes brighter in the near-IR than the limit of SNAPs Deep
Survey, the best follow-up capability will be offered
by the multi-object NIRSpec on JWST.

%


\begin{thebibliography}{}

\bibitem{}
Abadi, M.\ G., Moore, B., \& Bower, R.\ G.\ 1999, MNRAS, 308,
947

\bibitem{}
Abraham, R.\ G., et al.\ 1996, ApJ, 471, 694

\bibitem{}
Aldering, G., et al.\ 2004, PASP, submitted (astro-ph/0405232)

\bibitem{}
Baugh, C.\ M., Cole, S., Frenk, C.\ S., \& Lacey, C.\ G.\
1998, ApJ, 498, 504


\bibitem{}
Bell, E., et al.\ 2004, ApJ, 608, 752

\bibitem{}
Bender, R., Burstein, D., \& Faber, S.\ M.\ 1993, ApJ, 411, 153

\bibitem{}
Bender, R., et al.\ 1998, ApJ, 493, 529

\bibitem{}
Blakeslee, J.\ P., et al.\ 2003, ApJ, 596, L143

\bibitem{}
Bower, R.\ G., Lucey, J.\ R., \& Ellis, R.\ S.\ 1992, MNRAS, 254, 601

\bibitem{}
Butcher, H., \& Oemler, A., Jr.\ 1978, ApJ, 219, 18

\bibitem{}
Daddi, E., et al.\ 2003, ApJ, 588, 50

\bibitem{}
de Propris, R., Stanford, S.\ A., Eisenhardt, P.\ R., \& Dickinson,
M.\ 2003, ApJ, 598, 20

\bibitem{}
Diaferio, A., et al.\ 2001, MNRAS, 323, 999

\bibitem{}
Djorgovski, S., \& Davis, M.\ 1987, ApJ, 313, 59

\bibitem{}
Dressler, A.\ 1980, ApJ, 236, 351

\bibitem{}
Dressler, A., et al.\ 1997, ApJ, 490, 577

\bibitem{}
Dubinski, J.\ 1999, in ``Galaxy Dynamics'', ASP 182, Eds.\
D.\ Merritt, M.\ Valluri, \& J.\ A.\ Sellwood, p.\ 491

\bibitem{}
Ellingson, E., Lin, H., Yee, H.\ K.\ C., \& Carlberg, R.\ G.\
2001, ApJ, 547, 609

\bibitem{}
Ellis, R.\ S., et al.\ 1997, ApJ, 483, 582

\bibitem{}
Franx, M., et al.\ 2003, ApJ, 587, L79

\bibitem{}
F\"orster Schreiber, N.\ M., et al.\ 2004, ApJ, in press

\bibitem{}
Gladders, M.\ D., \& Yee, H.\ K.\ C.\ 2000, AJ, 120, 2148

\bibitem{}
Goto, T., et al.\ 2003, PASJ, 55, 757

\bibitem{}
Hoekstra, H., Franx, M., \& Kuijken, K.\ 2000, ApJ, 532, 88

\bibitem{}
Kelson, D., Illingworth, G., van Dokkum, P., \&
Franx, M.\ 2000, ApJ, 531, 184

\bibitem{}
Kelson, D., Illingworth, G., Franx, M., \& van Dokkum, P.\
2001, ApJ, 552, L17

\bibitem{}
Kenney, J.\ D.\ P., van Gorkom, J.\ H., Vollmer, B.\ 2004, AJ,
127, 3361

\bibitem{}
Kneib, J.-P., et al.\ 2003, ApJ, 598, 804

\bibitem{}
Kodama, T., Smail, I., Nakata, F., Okamura, S., \& Bower, R.\ G.\
2001, ApJ, 562, L9

\bibitem{}
Labb\'e, I., et al.\ 2003, AJ, 125, 1107

\bibitem{}
Meza, A., Navarro, J.\ F., Steinmetz, M., \& Eke, V.\ R.\ 2003,
ApJ, 590, 619

\bibitem{}
Moore, B., et al.\ 1996,
Nature, 379, 613

\bibitem{}
Pettini, M., et al.\ 2001, ApJ, 554, 981

\bibitem{}
Poggianti, B., et al.\ 1999, ApJ, 518, 576

\bibitem{}
Shapley, A., et al.\ 2001, ApJ, 562, 95


\bibitem{}
Smith, G., et al.\ 2004, ApJ, submitted (astro-ph/0403455)

\bibitem{}
Stanford, S.\ A., Eisenhardt, P.\ R., \& Dickinson, M.\ 1998,
ApJ, 492, 461

\bibitem{}
Steidel, C.\ C., et al.\ 1998, ApJ, 492, 428

\bibitem{}
Treu, T., et al.\ 2003, ApJ, 591, 53


\bibitem{}
van Dokkum, P.\ G., et al.\ 
1998a, ApJ, 500, 714

\bibitem{}
van Dokkum, P.\ G., Franx, M., Kelson, D.\ D., \& Illingworth, G.\ D.,
1998b, ApJ, 504, L17

\bibitem{}
van Dokkum, P.\ G., Franx, M., Fabricant, D., Kelson, D.\ D.,
\& Illingworth, G.\ D.\ 1999, ApJ, 520, L95

\bibitem{}
van Dokkum, P.\ G., \& Franx, M.\ 2001, ApJ, 553, 90

\bibitem{}
van Dokkum, P.\ G., \& Stanford, S.\ A.\ 2003, ApJ, 585, 78

\bibitem{}
van Dokkum, P.\ G., et al.\ 2003, ApJ, 587, L83

\bibitem{}
van Dokkum, P.\ G., et al.\ 2004, ApJ, in press (astro-ph/0404471)

\bibitem{}
Venemans, B.\ P., et al.\ 2002, ApJ, 569, L11

\bibitem{}
Wittman, D., Tyson, J.\ A., Margoniner, V.\ E., Cohen, J.\ G.,
\& Dell'Antonio, I.\ P.\ 2001, ApJ, 557, L89

\bibitem{}
Wuyts, S., van Dokkum, P.\ G., Kelson, D.\ D., Franx, M., \& Illingworth,
G.\ D.\ 2004, ApJ, 605, 677

\end{thebibliography}
\end{document}